\documentclass[12pt]{article}
\begin{document}
\begin{titlepage}
\begin{flushright}
hep-th/0210067 \\
TIT/HEP-486 \\
October, 2002 \\
\end{flushright}
\vspace{0.5cm}
\begin{center}
{\Large \bf
On the Moduli Space of Noncommutative Multi-solitons at Finite $\theta$
}
\lineskip .75em
\vskip2.5cm
{\large Takeo Araki } and {\large Katsushi Ito}
\vskip 1.5em
{\large\it Department of Physics\\
Tokyo Institute of Technology\\
Tokyo, 152-8551, Japan}
\vskip 3.5em
\end{center}
\vskip3cm
\begin{abstract}
We study the finite $\theta$ correction to the metric of the moduli
space of noncommutative multi-solitons in scalar field theory in (2+1)
dimensions.
By solving the equation of motion up to order
$O(\theta^{-2})$ explicitly,
we show that the multi-soliton solution must have the
same center for a generic potential term. We examine the condition
that the multi-centered configurations are allowed.
Under this condition, we calculate the finite $\theta$ correction to the
metric of the moduli space of multi-solitons and argue
the possibility of the non right-angle scattering of two solitons.
We also obtain the potential between two solitons. 

\end{abstract}
\end{titlepage}
\baselineskip=0.7cm

Solitons in field theories on noncommutative spacetime are useful
for studying non-perturbative effects.
These play also an important role in
 string theory and condensed matter physics(see
\cite{HaDoNe} for reviews).
In particular, the noncommutative soliton in $2+1$ dimensional
scalar field theory \cite{GMS} provides an interesting and nontrivial
example since it does not exist in scalar field theory on
commutative spacetime.

Various aspects of noncommutative scalar solitons have been studied
\cite{GHS,LRU,HLRU,AI,DJ,Mat,noncomm}.
In ref. \cite{GHS}, the multi-soliton solutions and their moduli space are
studied in the limit of large noncommutative parameter $\theta$.
The geodesic in the moduli space of two solitons describes
the scattering of the soliton in the adiabatic approximation\cite{Ma}.
In refs. \cite{LRU,HLRU,AI}, it is shown that the right-angle scattering
of solitons
occurs at the head-on collision.
The scattering is also studied for various noncommutative solitons
\cite{sca}.

It is an interesting problem whether noncommutative solitons in scalar field
theory exist at finite $\theta$.
In this case, we need to consider the attractive or repulsive force
between solitons.
Recently, Durhuus and Jonsson pointed out that there are no
multi-soliton solutions which interpolate smoothly between $n$
overlapping solitons and $n$ solitons with an infinite separation
at the lowest order perturbation in $\theta^{-1}$\cite{DJ}.
Therefore multi-solitons at finite $\theta$ are in general unstable and
decay into infinitely separate or overlapping solitons.
This  heavily depends on the shape of potential term.
In fact, for the potential $V(\phi)$ with
$1/V''(0)+1/V''(\lambda)=0$, where $\phi=0$ and $\lambda$ are two critical
points of $V(\phi)$, energies for the above two configurations agree
with each other.
Hence the moduli space approximation looks still good for such a
potential.

In this paper,
we study the $\theta$ correction to the noncommutative scalar soliton.
We solve the static
equation
of motion explicitly in scalar field theory
up to the order
 $O(\theta^{-2})$.
We examine some consistency conditions
which appears in the process of solving the equation of motion.
{}From these conditions,
we show
that  noncommutative multi-solitons must have the
same center for the potential with $1/V''(0)+1/V''(\lambda)\neq 0$.
On the other hand, multi-centered
configurations are allowed for the  potential with
$1/V''(0)+1/V''(\lambda)= 0$.
Using the perturbed solutions, we may calculate the finite
$\theta$ correction
to the metric of the moduli
space for the  multi-solitons.
We argue the possibility of the
non right-angle scattering for scattering of two solitons for finite
$\theta$. This would provide interesting physics  in contrast
with the right-angle scattering in the large $\theta$ limit.
We also study the force between solitons by calculating the potential
and discuss existence of a static multi-soliton solution
at finite $\theta$.

We consider two-dimensional noncommutative
space coordinates $(\hat{x},\hat{y})$  satisfying
$
[\hat{x},\hat{y}]=i\theta.
$
We rescale the coordinates by the factor $\sqrt{\theta}$ such that
$
[\hat{x},\hat{y}]=i
$.
Introducing the harmonic oscillators
$a={1\over\sqrt{2}}(\hat{x}-i\hat{y})$ and
$a^{\dagger}={1\over\sqrt{2}}(\hat{x}+i\hat{y})$,
the action of the real scalar field theory is
\begin{equation}
 S=\int dt 2\pi {\rm Tr}\left\{ \theta (\partial_{t}\phi)^2
+[a,\phi] [\phi,a^{\dagger}]+\theta V(\phi)
\right\},
\end{equation}
where the trace is taken over the Fock space ${\cal H}$ of the harmonic
oscillator.
Here we assume that the potential term $V(\phi)$ has critical points
at $\phi=0$ and $\lambda$.
In the large $\theta$ limit, the second term in the action drops out.
The static equation of motion becomes $V'(\phi)=0$. Then
it admits the soliton solution of the form\cite{GMS}
\begin{equation}
 \phi_{0}=\lambda P,
\label{eq:sol0}
\end{equation}
where $P$ is a projection operator $P^2=P$.
The multi-soliton solution \cite{GHS} is constructed by using
the coherent state
$|z\rangle\equiv e^{a^{\dagger}z}|0\rangle$.
The $n$ level one solitons centered at points $z_{1},\cdots, z_{n}$ are
given by
\begin{equation}
P=|z_{i}\rangle h^{ij} \langle z_{j}|, \quad
h_{ij}=\langle z_{i} | z_{j} \rangle,
\label{eq:proj1}
\end{equation}
where $h^{ij}$ is inverse matrix of $h_{ij}$.
One may also construct the level $k$ soliton centered at $z$ in terms of
basis
$|\psi_{i}\rangle=(a^{\dagger})^{i}|z\rangle$ ($i=0,\cdots, k-1$):
\begin{equation}
 P=|\psi_{i}\rangle H^{ij} \langle \psi_{j}|, \quad
H_{ij}=\langle \psi_{i} | \psi_{j} \rangle.
\label{eq:proj2}
\end{equation}
Here $H^{ij}$ satisfies $H^{ij}H_{jk}=\delta^{i}_{k}$.
For large but finite $\theta$, we must consider both the second and the
third
terms in the action. The energy functional
\begin{equation}
 E=2\pi {\rm Tr}\left(\theta V(\phi)+[a,\phi] [\phi,a^{\dagger}] \right)
\label{eq:ener}
\end{equation}
leads to the equation of motion:
\begin{equation}
 2[a^{\dagger},[a,\phi]]+\theta V'(\phi)=0.
\label{eq:eom1}
\end{equation}
We want to obtain the solution of (\ref{eq:eom1}) of the
form
\begin{equation}
 \phi=\phi_{0}+{1\over\theta}\phi_{1}+{1\over\theta^2}\phi_{2}+\cdots .
\label{eq:sol1}
\end{equation}
In ref. \cite{HLRU},
$\phi_{1}$ has been constructed.
We will calculate the $\phi_{2}$-term.

Substituting (\ref{eq:sol1}) into (\ref{eq:eom1}), we obtain
\begin{equation}
 2\sum_{r=0}^{\infty} \theta^{-r} [a^{\dagger},[a,\phi_{r}]]
+\sum_{r_{0}\geq r_{1}\geq \geq 0}
\theta^{1-\sum_{i}r_{i}} \left(
V^{(r_{0}+1)}(\phi_{0})
{\phi_{1}\over (r_{0}-r_{1})!} {\phi_{2}\over (r_{1}-r_{2})!}
\cdots
\right)_{S}=0,
\end{equation}
where $({\cal O}_{1}\cdots {\cal O}_{n})_{S}={1\over n!}
\sum_{\sigma\in S_{n}}{\cal O}_{\sigma(1)}\cdots {\cal O}_{\sigma(n)}$
denotes the symmetrized sum of the product
${\cal O}_{1}\cdots {\cal O}_{n}$.
$S_{n}$ is the permutation group of order $n$.
Taking the coefficient of $\theta^{-r}$ of the equation of motion,
we have a series of equations for $\phi_{r}$'s.
The first three equations become
\begin{eqnarray}
&& V^{(1)}(\phi_{0})=0,\nonumber\\
&& 2[a^{\dagger},[a,\phi_{0}]]+(V^{(2)}(\phi_0)\phi_1)_{S}=0,\nonumber\\
&& 2[a^{\dagger},[a,\phi_{1}]]+{1\over2!}(V^{(3)}(\phi_0)\phi_1^2)_{S}
+(V^{(2)}(\phi_0)\phi_2)_{S}=0.
\label{eq:eom2}
\end{eqnarray}
The first equation in (\ref{eq:eom2}) has a solution $\phi_{0}=\lambda
P$.
In order to solve the second and the third equations, it is convenient to
use the formulas for an operator $A$:
\begin{eqnarray}
 (V^{(n)}(\phi_{0})A)_{S}&=& V^{(n)}(\lambda) PAP+V^{(n)}(0)QAQ\nonumber\\
&&
+{1\over\lambda}\left(V^{(n-1)}(\lambda)-V^{(n-1)}(0) \right)
(PAQ+QAP),\\
{1\over2!}  (V^{(n)}(\phi_{0})A^2)_{S}&=&
{1\over2!} V^{(n)}(\lambda)PAPAP+{1\over2!}V^{(n)}(0)QAQAQ\nonumber\\
&&\hspace{-2cm}
-{1\over\lambda^2} \left\{
V^{(n-2)}(\lambda)-V^{(n-2)}(0)-\lambda V^{(n-1)} (\lambda)
\right\} (PAPAQ+PAQAP+QAPAP)\nonumber\\
&&\hspace{-2cm}
+{1\over\lambda^2} \left\{
V^{(n-2)}(\lambda)-V^{(n-2)}(0)-\lambda V^{(n-1)} (0)
\right\} (PAQAQ+QAPAQ+QAQAP)
\nonumber\\
\label{eq:ident1}
\end{eqnarray}
where $Q=1-P$.
Since $V^{(1)}(0)=V^{(1)}(\lambda)=0$, we have
\begin{equation}
 (V^{(2)}(\phi_{0})A)_{S}=V^{(2)}(\lambda) PAP+V^{(2)}(0)QAQ.
\end{equation}
Since the r.h.s. of this equation does not include the off-diagonal
parts $PAQ$ and $QAP$,
each equation in (\ref{eq:eom2}) gives
further constraints on  the structure of the solution.
For example, from the off-diagonal parts $P(\dots)Q$ and $Q(\dots)P$
of the second equation in (\ref{eq:eom2}),
we can see that $\phi_{0}$ must satisfy
\begin{eqnarray}
 P [a^{\dagger},[a,\phi_{0}]] Q=0, \quad
 Q [a^{\dagger},[a,\phi_{0}]] P=0.
\label{eq:condition1}
\end{eqnarray}
These equations are satisfied if the projection operator
$P$ obeys the equation $(1-P)aP=0$, which means
$a P{\cal H}\subset P{\cal H}$.
When $P$ is constructed
from the coherent states of the form (\ref{eq:proj1}),
this condition is satisfied.
In this case, $\phi_1$ is regarded as the leading correction to $\phi_0$
and given by
\begin{equation}
 \phi_{1}=-{2\lambda\over V^{(2)}(\lambda)} P[a^{\dagger},[a,P]]P
-{2\lambda\over V^{(2)}(0)} Q[a^{\dagger},[a,P]]Q
+PX_{1}Q+QX_{1}^{\dagger}P,
\label{eq:phi1a}
\end{equation}
where $X_{1}$ is an arbitrary operator.
Since $P$ is the projection operator of the form (\ref{eq:proj1}),
we can put $QaP=0$ and (\ref{eq:phi1a}) may be simplified to
\begin{equation}
 \phi_{1}=-{2\lambda\over V^{(2)}(\lambda)} PaQa^{\dagger}P
+{2\lambda\over V^{(2)}(0)} Qa^{\dagger}PaQ
+PX_{1}Q+QX_{1}^{\dagger}P.
\label{eq:phi1b}
\end{equation}

The second order correction $\phi_{2}$ can be solved in a similar way.
The consistency condition from the third equation in (\ref{eq:eom2}) yields
\begin{equation}
2P[a^{\dagger},[a,\phi_{1}]]Q+
{1\over\lambda}(V^{(2)}(\lambda)P\phi_{1}P\phi_{1}Q-V^{(2)}(0)
P\phi_{1}Q\phi_{1}Q)=0
\label{eq:const2}
\end{equation}
where we have used the second equation in (\ref{eq:ident1}).
Using $QaP=0$ and (\ref{eq:phi1b}), (\ref{eq:const2}) turns out to be
\begin{eqnarray}
F_{2}&\equiv& 4\lambda\left(
{1\over V^{(2)}(\lambda) }+{1\over V^{(2)}(0)}\right)
(Pa^{\dagger}PaQa^{\dagger}PaQ-PaQa^{\dagger}Pa^{\dagger}PaQ)
\nonumber\\
&& +2\left(
PaPa^{\dagger}PX_{1}Q-PaPX_{1}Qa^{\dagger}Q
-Pa^{\dagger}PX_{1}QaQ+PX_{1}Qa^{\dagger}QaQ
\right)=0.
\label{eq:cons2}
\end{eqnarray}
One may ask whether this consistency condition provides further
constraints on $P$ and $PX_{1}Q$.

When $1/V^{(2)}(\lambda)+1/V^{(2)}(0)\neq 0$
the first term
in (\ref{eq:cons2}) is not zero for a multi-soliton projection operator $P$.
In fact, we may evaluate the matrix elements of the operator $F_{2}$
on the basis
$\{ |z_{1}\rangle,\cdots,|z_{n}\rangle, Q|z\rangle
\ | z\in{\bf C}, z\neq z_{i} \}$ which spans ${\cal H}$.
Non-trivial elements are given by
\begin{equation}
\langle z_{i}|F_{2}Q|z\rangle=A
\langle z_{i}|a Q a^{\dagger}|z_{p}\rangle
h^{pq} (\bar{z}_{i}-\bar{z}_{q}) \langle z_{q}| aQ|z\rangle-2
N_{i}^{q} \langle z_{q}|[PX_{1}Q,a^{\dagger}]Q|z\rangle
\end{equation}
where
$A=4\lambda\left(
{1\over V^{(2)}(\lambda) }+{1\over V^{(2)}(0)}\right)$ and
$N_{i}^{q}=h_{ip}(z-z_{p})h^{pq}$.
{}From this formula, we can see that the first term is zero
for the level $k$ soliton projection operator (\ref{eq:proj2})
but non-zero for the multi-soliton case (\ref{eq:proj1}).
In the latter case, we are not able to find $PX_{1}Q$ so that
the non-zero first term is canceled by the second one.
In the former case, on the other hand,
one may choose $PX_{1}Q=0$ in order to make $F_{2}$ zero.
These mean the following:
if we take into account the $O(\theta^{-2})$-correction,
the multi-soliton configuration
cannot be a solution of the equation of
motion for generic potential with $1/V^{(2)}(\lambda)+1/V^{(2)}(0)\neq
0$, but the single level $k$ soliton configuration
can.
This seems to be consistent with the argument based on the
evaluation of the energy functional in \cite{GHS,HLRU,DJ}.

In the case of  $1/V^{(2)}(\lambda)+1/V^{(2)}(0)=0$, however, the
consistency
condition is satisfied for a multi-soliton projection operator $P$ and
$X_{1}$
with $PX_{1}Q=0$.
Hence we assume $PX_{1}Q=0$ from now on.
We find that $\phi_{2}$ is given by
\begin{eqnarray}
 \phi_{2}&=& -{1\over V^{(2)}(\lambda)}
\left(
2P[a^{\dagger},[a,\phi_{1}]]P+{1\over2!}V^{(3)}(\lambda)P\phi_{1}P\phi_{1}P
+{1\over\lambda}V^{(2)}(\lambda)P\phi_{1}Q\phi_{1}P
\right)\nonumber\\
&&
 -{1\over V^{(2)}(0)}
\left(
2Q[a^{\dagger},[a,\phi_{1}]]Q+{1\over2!}V^{(3)}(0)Q\phi_{1}Q\phi_{1}Q
-{1\over\lambda}V^{(2)}(0)Q\phi_{1}P\phi_{1}Q
\right)\nonumber\\
&&+PX_{2}Q+QX_{2}^{\dagger}P
\label{eq:phi2a}
\end{eqnarray}
where $X_{2}$ is an arbitrary operator.
We will assume $X_2=0$ for simplicity.
Substituting (\ref{eq:phi1b}) into (\ref{eq:phi2a}) and using $QaP=0$,
one may obtain an explicit formula for $\phi_2$:
\begin{eqnarray}
 \phi_2&=&
\lambda\left\{
 v^2(P aPa^\dagger P aQa^\dagger P
 - P aP aQa^\dagger Pa^\dagger P - Pa^\dagger P aQa^\dagger P aP
 + P aQa^\dagger P aPa^\dagger P \right.\nonumber\\
&&{} \left.\quad - P aQa^\dagger P)
 +\left(2v^2-vw+\frac12 C_1
% + {\lambda\over2!}{v^3\over2}V^{(3)}(\lambda)
 \right)\,
 P aQa^\dagger P aQa^\dagger P\right\}\nonumber\\
&&{}-\lambda\left\{
 w^2(Qa^\dagger P aQ + Qa^\dagger Q aQa^\dagger P aQ
 - Q aQa^\dagger P aQa^\dagger Q
 - Qa^\dagger Qa^\dagger P aQ aQ \right.\nonumber\\
&&{} \left. \quad + Qa^\dagger P aQa^\dagger Q aQ)
 +\left(2w^2-vw+\frac12 C_2
% +{\lambda\over2!}{w^3\over2}V^{(3)}(0)
 \right)\,
 Qa^\dagger P aQa^\dagger P aQ \right\},
\label{eq:phi2b}
\end{eqnarray}
where
\begin{eqnarray}
&&
v=-2/V^{(2)}(\lambda), \quad
w=2/V^{(2)}(0),
\label{eq:coeff1}\\
&&
C_{1}={\lambda\over 2!}v^3V^{(3)}(\lambda),\quad
C_{2}={\lambda\over 2!}w^3 V^{(3)}(0).
\label{eq:coeff2}
\end{eqnarray}
We have written the equation (\ref{eq:phi2b}) for generic $v$ and $w$.
But the condition $1/V^{(2)}(\lambda)+1/V^{(2)}(0)=0$ implies $v=w$,
so we should impose this condition on (\ref{eq:phi2b}).

Obtaining the solution of the equation of motion, we may calculate the
metric of the moduli space and its $\theta^{-1}$-correction.
In the case of $\theta=\infty$, the moduli space of $n$ level one soliton
solution are parameterized by the coordinates $z_{i},\bar{z}_{i}$
($i=1,\cdots,n $)
of the level one solitons.
Using the adiabatic approximation \cite{Ma}, the metric of the moduli space
is
determined  by the action of the particles:
\begin{equation}
 S=\int dt \left(g_{i\bar{\jmath}}
 {d z_{i}\over dt} {d \bar{z}_{j}\over dt}
+g_{ij} {d z_{i}\over dt} {d z_{j}\over dt}
+g_{\bar{\imath}\bar{\jmath}}
 {d \bar{z}_{i}\over dt} {d \bar{z}_{j}\over dt} \right).
\end{equation}
Here the metric is given by the formula
\begin{eqnarray}
 g_{i\bar{\jmath}}&=& {1\over\lambda^2}
{\rm Tr}\partial_{i}\phi \bar{\partial}_{j}\phi,
\nonumber\\
g_{ij}&=& {1\over\lambda^2}
{\rm Tr}\partial_{i}\phi \partial_{j}\phi,
\nonumber\\
g_{\bar{\imath}\bar{\jmath}}&=&
{1\over\lambda^2}
{\rm Tr}\bar{\partial}_{i}\phi \bar{\partial}_{j}\phi
\end{eqnarray}
where $\partial_{i}={\partial\over \partial z_{i}}$ and
$\bar{\partial}_{i}={\partial\over \partial \bar{z}_{i}}$.
{}From (\ref{eq:sol1}), the metric can be expanded as
\begin{eqnarray}
g_{i\bar{\jmath}}&=&
g^{(0)}_{i\bar{\jmath}}+{1\over \theta}g^{(1)}_{i\bar{\jmath}}
+{1\over \theta^2}g^{(2)}_{i\bar{\jmath}}+\cdots ,
\nonumber\\
g_{ij}&=& g^{(0)}_{ij}+{1\over \theta}g^{(1)}_{ij}
+{1\over \theta^2}g^{(2)}_{ij}+\cdots ,
\nonumber\\
g_{\bar{\imath}\bar{\jmath}}&=&
g^{(0)}_{\bar{\imath}\bar{\jmath}}
+{1\over \theta}g^{(1)}_{\bar{\imath}\bar{\jmath}}
+{1\over \theta^2}g^{(2)}_{\bar{\imath}\bar{\jmath}}+\cdots .
\end{eqnarray}
In the large  $\theta$ limit, the metric is shown to satisfy
the K\"ahler condition\cite{GHS}:
\begin{equation}
\partial_{k} g^{(0)}_{i\bar{\jmath}}
=\partial_{i}g^{(0)}_{k\bar{\jmath}},
\quad  g^{(0)}_{\bar{\imath}\bar{\jmath}}=g^{(0)}_{ij}=0.
\end{equation}
The K\"ahler structure comes from the Grassmannian structure of the
moduli space \cite{Mat} which can be regarded as the space of symmetric
products of ${\bf R}^{2}$ \cite{GHS}.

We begin with  the first
order correction to the metric for the multi-solitons.
The first order correction to the metric for two solitons has been
calculated in \cite{HLRU}.
In order to simplify the formula, we use the following differential
operators which take values in the space of operators on the Hilbert
space ${\cal H}$:
\begin{equation}
\partial P=\partial_{i}P d z_{i},\quad
\bar{\partial}P=\bar{\partial}_{i}P d\bar{z}_{i}.
\end{equation}
Using $P=|z_{i}\rangle h^{ij} \langle z_{j}|$, we have
\begin{equation}
 \partial_{i}P=Q a^{\dagger} |z_{i}\rangle h^{il} \langle z_{l}|,
\quad  \bar{\partial}_{i}P=|z_{l}\rangle  h^{li} \langle z_{i}|a Q.
\end{equation}
Then it is shown that the K\"ahler form $K$ takes the form \cite{GHS}
\begin{equation}
 K={i\over 2}{\rm Tr}\partial P\wedge \bar{\partial}P
={i\over 2}g^{(0)}_{i\bar{\jmath}}dz_{i}\wedge d\bar{z}_{j}
\end{equation}
with
\begin{equation}
 g^{(0)}_{i\bar{\jmath}}=h^{ij} M_{ji}
\label{eq:g0a}
\end{equation}
and
\begin{equation}
 M_{ji}\equiv \langle z_{j}|aQa^{\dagger}| z_{i}\rangle
=\bar{z}_{j}h_{ji}z_{i}+h_{ji}-h_{jl}z_{l}h^{lm}\bar{z}_{m}h_{mi}.
\label{eq:M}
\end{equation}
The first order correction to the metric $g^{(1)}$ is
\begin{equation}
 g^{(1)}_{i\bar{\jmath}}={1\over\lambda^2}
{\rm Tr}[\partial_{i}\phi_{0}\partial_{\bar{\jmath}}\phi_{1}
+\partial_{i}\phi_{1}\partial_{\bar{\jmath}}\phi_{0}]
\end{equation}
and similar expressions for $g^{(1)}_{ij}$ and
$g^{(1)}_{\bar{\imath}\bar{\jmath}}$.
Using (\ref{eq:phi1b}), we have
\begin{eqnarray}
 g^{(1)}_{i\bar{\jmath}}&=&2vh^{il}M_{lm}h^{mj}M_{ji}
-2w h^{ij} M_{jm} h^{ml} M_{li},\nonumber\\
g^{(1)}_{ij}&=& g^{(1)}_{\bar{\imath}\bar{\jmath}}=0,
\end{eqnarray}
where $v$ and $w$ have been defined in (\ref{eq:coeff1}).

Let us check whether the metric including
$O(\theta^{-1})$-correction
satisfies the K\"ahler condition.
After some computations, we find
\begin{equation}
\partial_{k}g^{(1)}_{i\bar{\jmath}}-\partial_{i} g^{(1)}_{k\bar{\jmath}}
=(v-w)\left( h^{il}(\bar{z}_{l}-\bar{z}_{j})M_{lk}h^{kj}M_{ji}
-(i\leftrightarrow k)
\right) .
\end{equation}
These quantities vanish if $v=w$  or the second factor becomes zero.
In the case of $v=w$ i.e. $1/V^{(2)}(\lambda)+1/V^{(2)}(0)=0$,
the metric $g^{(0)}+{1\over\theta}g^{(1)}$ is K\"ahler
for any multi-soliton configuration.
In the case of $v\neq w$ i.e. $1/V^{(2)}(\lambda)+1/V^{(2)}(0)\neq 0$,
on the other hand,
the second factor does not vanish for the multi-soliton configuration,
thus the metric $g^{(0)}+{1\over\theta}g^{(1)}$ is not K\"ahler.
But it is still a complex manifold because of the property
$g_{ij}=g_{\bar{\imath}\bar{\jmath}}=0$.
For the single level $k$ soliton configuration
and the two soliton configuration,
the K\"ahler structure is trivial,
even if $v\neq w$.

We next consider the second order correction to the metric.
As we have discussed, we need to impose the condition $v=w$
in order that a multi-soliton configuration is allowed
as the solution of the equation of motion up to $O(\theta^{-2})$.
The metric $g^{(2)}$ is obtained from
\begin{equation}
 g^{(2)}_{i\bar{\jmath}}={1\over\lambda^2}{\rm Tr}
[\partial_{i}\phi_{0}\bar{\partial}_{j}\phi_{2}
+\partial_{i}\phi_{1}\bar{\partial}_{j}\phi_{1}
+\partial_{i}\phi_{2}\bar{\partial}_{j}\phi_{0}]
\end{equation}
etc.
Using (\ref{eq:phi1b}) and (\ref{eq:phi2b}), we get
\begin{eqnarray}
 g^{(2)}_{i\bar{\jmath}}&=&
\left(
5v^2-2vw+C_{1}
\right) h^{il}M_{lm}h^{mn} M_{np} h^{pj}M_{ji}
\nonumber\\
&& +
\left(
5w^2-2vw+C_{2}
\right) h^{ij}M_{jm}h^{mn}M_{np}h^{pl}M_{li}
\nonumber\\
&& +
(v^2-2w^2) \left\{ (\bar{z}_{n}-\bar{z}_{j})(z_{m}-z_{j})-1\right\}
h^{ij} M_{jm}h^{mn}M_{ni}
\nonumber\\
&&
+(w^2-2v^2) \left\{ (\bar{z}_{l}-\bar{z}_{j})(z_{m}-z_{j})+1\right\}
h^{il} M_{lm}h^{mj}M_{ji}
\nonumber\\
&& +
(v-w)^2 h^{il}M_{lm}h^{mj}M_{jn}h^{np}M_{pi}
\label{eq:met2a}
\end{eqnarray}
and
\begin{equation}
 g^{(2)}_{ij}=-(v^2+w^2)
(\bar{z}_{l}-\bar{z}_{m})^2h^{jl}M_{li}h^{im}M_{mj}.
\label{eq:met2b}
\end{equation}
In (\ref{eq:met2a}) and (\ref{eq:met2b}), we have written the formulas
for generic $v$ and $w$. But as we have noted, we must put $v=w$.
The condition $v=w$ implies that
the K\"ahler condition holds at the first order,
as mentioned before.
However, taking into account the second order correction,
the metric is neither K\"ahler nor complex.

We give an example of the metric for two level one solitons
($(1,1)$-system).
The metric  of two-soliton solution and its behavior
around the origin are particularly interesting because the geodesic
under the metric describes the scattering of two noncommutative
solitons.
The effect of the first order corrections has been analyzed in
ref. \cite{HLRU}.
The first order correction does not change the structure of the
right-angle scattering property.
Let us examine the effect of the second order perturbation.
In  the center of mass system, i.e.
$z_{1}=-z_{2}=y/2$, we get
\begin{eqnarray}
 g^{(0)}_{y\bar{y}}
&=&
\coth t -{t\over \sinh^2 t},
\\
g^{(1)}_{y\bar{y}}&=&
2(v-w)\left(
\coth t -{t\over \sinh^2 t}
\right)
+2(v+w) {t\over \sinh^2 t} (1-t \coth t)
\end{eqnarray}
and
\begin{eqnarray}
g^{(2)}_{y\bar{y}}&=&
{1\over \sinh^4 t}
\Bigl[
v^2\{
-t^3(2\cosh^2 t+3) -3t^2\sinh t\cosh t+5t \sinh^2 t+3\sinh^3 t\cosh t
\}\nonumber\\
&& +vw \{
2t^3 -2t^2 \sinh t\cosh t +6 t\sinh^2 t-6\sinh^3 t\cosh t
\} \nonumber\\
&& +w^2\{
-t^3(2\cosh^2 t+3) +17 t^2\sinh t\cosh t-21t\sinh^2 t+9\sinh^3 t\cosh t
\}\nonumber\\
&& +C_{1} \{
-t^3-t^2\sinh t\cosh t +t\sinh^2 t+\sinh^3 t\cosh t
\}\nonumber\\
&&+C_{2} \{
-t^3+3t^2\sinh t\cosh t-3 t \sinh^2 t+\sinh^3 t\cosh t
\}
\Bigr],
\\
g^{(2)}_{yy}&=& {(v^2+w^2)\bar{y}^2\over 2\sinh^2 t} (1-t\cosh t)^2.
\end{eqnarray}
Here $t=|y|^2/2$ and $v=w$.

A particularly interesting fact from the above results is that
each $g^{(i)}$ have the same conical singularity at the origin $t=0$.
This can be explicitly seen when we transform the metric into the K\"ahler
form.
We change the coordinates from $(y,\bar{y})$ to $(y',\bar{y}')$ by
\begin{equation}
 y'=y+{1\over\theta^2} \delta y(y,\bar{y}),\quad
 \bar{y}'=\bar{y}+{1\over\theta^2} \delta \bar{y}(y,\bar{y}),
\end{equation}
where $\delta y$ satisfies
\begin{equation}
 {\partial \delta y\over \partial y}={1\over 2}
{g^{(2)}_{yy}\over g^{(0)}_{y\bar{y}}}.
\end{equation}
The solution of the above equation takes the form
\begin{equation}
 \delta y={1\over 2} H(|y|^2)y+\mbox{holomorphic function of $y$}
\end{equation}
where $H(|y|^2)$ satisfies
\begin{equation}
 {d H(|y|^2)\over d |y|^2}={g^{(2)}_{yy}\over g^{(0)}_{y\bar{y}}|y|^2}.
\end{equation}
After change of the coordinates, the metric for finite $\theta$
becomes
$
 ds^2=g_{y'\bar{y}'}dy' d\bar{y}'
$
where
\begin{equation}
 g_{y'\bar{y}'}=g^{(0)}_{y'\bar{y}'}+{1\over\theta} g^{(1)}_{y'\bar{y}'}
+{1\over \theta^2} g'{}^{(2)}_{y'\bar{y}'}+\cdots
\end{equation}
and
\begin{equation}
  g'{}^{(2)}_{y'\bar{y}'}=g^{(2)}_{y'\bar{y}'}-g^{(2)}_{y'y'} {|y'|^2
\over \bar{y}'{}^2}-g^{(0)}_{y'\bar{y}'} H(|y'|^2).
\end{equation}
Near $|y'|^2=0$ we have
\begin{eqnarray}
 g^{(0)}_{y'\bar{y}'}&=& {2\over3}{|y'|^2\over2}+O(|y'|^6),\nonumber\\
g^{(1)}_{y'\bar{y}'}&=& \Bigl(
{2\over3}v-2w\Bigr){|y'|^2\over2}+O(|y'|^6),\nonumber\\
  g'{}^{(2)}_{y'\bar{y}'}&=&\Bigl(
{2\over 3}v^2-{16\over 3}vw+{34\over3}w^2+{2\over3}C_{1}+2C_{2}
\Bigr){|y'|^2\over2}+O(|y'|^6).
\end{eqnarray}
{}From the above results one might expect the metric for $(1,1)$-system
for finite $\theta$ takes the form
\begin{equation}
 ds^2=f(r,\theta)(dr^2+r^2d\varphi^2)
\end{equation}
in the polar coordinate $(r,\varphi)$.
The conformal factor $f(r,\theta)$ has an expansion
\begin{equation}
f(r,\theta)=f_{1}(\theta)r^2+f_{3}(\theta)r^6+\cdots
\label{eq:expa}
\end{equation}
If $\theta$ satisfies $f_{1}(\theta)=0$, the right-angle scattering does
not occur at the head-on collision. In this case, the leading term in
(\ref{eq:expa}) is $r^6$ term and
the ${\pi/6}$ scattering occurs instead.
Indeed, due to symmetry,  it is very hard to think about
non-right angle scattering for the same type of solitons.
In order to see whether this phenomena really happen,
we need to obtain the exact solution of
the equation of motion for finite $\theta$.

For the study of dynamical aspects of the solitons at finite $\theta$,
we also consider the potential $U(|y|)$ for two solitons, which is
obtained by substituting the solution (\ref{eq:sol1}) into the energy
functional (\ref{eq:ener}) \cite{GHS,HLRU}.
The energy functional $E$ can be expanded as
\begin{eqnarray}
E&=&2\pi\theta{\rm Tr}\left[
 \sum_{r=0}^\infty\theta^{-r-1}\sum_{m=0}^r
 [a,\phi_{r-m}][\phi_m,a^\dagger]
 \right.\nonumber\\
&&\left.{}
 +\sum_{r_0\ge r_1\ge\dots\ge0}^{\infty}
 \theta^{-\sum_{i=0}^\infty r_i}
 \left(V^{(r_0)}(\phi_0){\phi_1^{r_0-r_1}\over(r_0-r_1)!}
 {\phi_2^{r_1-r_2}\over(r_1-r_2)!}\dots\right)_S
 \right]
\nonumber\\
&=&2\pi\theta \left(E_0+{1\over\theta}E_1
 +{1\over\theta^{2}}E_2+{1\over\theta^{3}}E_3+\dots\right),
\label{eq:expanded_e}
\end{eqnarray}
where
\begin{eqnarray}
&&E_0[\phi_0]={\rm Tr} V(\phi_0), \nonumber\\
&&E_1[\phi_0]={\rm Tr}[a^\dagger ,[a,\phi_0]]\phi_0, \nonumber\\
&&E_2[\phi_0,\phi_1]
 ={\rm Tr}\left[2[a^\dagger ,[a,\phi_0]]\phi_1
 +(V^{(2)}(\phi_0){\phi_1^2\over 2!})_S\right], \nonumber\\
&&E_3[\phi_0,\phi_1]\nonumber\\
&& ={\rm Tr}\left[
 \left\{2[a^\dagger ,[a,\phi_0]]
 +(V^{(2)}(\phi_0)\phi_1)_S\right\}\phi_2
 +[a^\dagger ,[a,\phi_1]]\phi_1
 +(V^{(3)}(\phi_0){\phi_1^3\over 3!})_S\right]
\label{eq:energies}
\end{eqnarray}
and so on.
Here we have already used $V^{(1)}(\lambda)=V^{(1)}(0)=0$.
For a rank $k$ projection operator $P$, $E_0=kV(\lambda)$.
Further assuming $QaP=0$, we have $E_1=\lambda^2 k$.
These are constants and are not included in the potential $U$.
For $(1,1)$-system, $U$ takes the form
\begin{equation}
 U={1\over\theta} U^{(1)}+{1\over \theta^2} U^{(2)}+\cdots
\end{equation}
where
\begin{eqnarray}
 U^{(1)}&=& 2(v-w){t^2\over\sinh^2 t }\nonumber\\
 U^{(2)}&=& -8 (v^2-w^2) {t^2(1-t\coth t)\over \sinh^2 t}
-4 (13v^2-7w^2-{1\over2}C_{1}-{1\over2}C_{2}) {t^2\over \sinh^2 t}.
\end{eqnarray}
$U^{(1)}$ is obtained in \cite{HLRU},
which comes from
\begin{equation}
E_2[\phi_0,\phi_1]
 =-{(2\lambda)^2\over2!}\left({1\over V^{(2)}(\lambda)}
 +{1\over V^{(2)}(0)}\right)
 {\rm Tr} ( P  a Q a^\dagger  P )^2 .
\label{eq:e2b}
\end{equation}
$U^{(2)}$ comes from
\begin{eqnarray}
E_3[\phi_0,\phi_1]
 &=&\lambda^2{\rm Tr}\left[
 \left\{{1\over3}C_1
 +{1\over3}C_2
 + 2v^2 - 2vw +2w^2 \right\}( P  a Q a^\dagger  P )^3
 \right.\nonumber\\
&&{}
 + 2(v^2-w^2) Q  a Q a^\dagger P
 ( P a^\dagger P  a Q a^\dagger P  a Q
 -  P  a Q a^\dagger P a^\dagger P  a Q  )\nonumber\\
&&{} \left.- (v^2-w^2) ( P  a Q a^\dagger  P )^2 \right].
\label{eq:e3b}
\end{eqnarray}
Here we have dropped the constant terms coming from $E_2$ and $E_3$.
The formula (\ref{eq:e3b}) is valid for generic $v$ and $w$,
as long as $\phi_0$ and $\phi_1$ are
the consistent solution of the equation of motion up to $O(\theta^{-1})$.
Of course we should take $v=w$ for $(1,1)$-system.
In this case, $U^{(1)}$ and
the first term in $U^{(2)}$ become zero.
Assuming that
the coefficient of the second term in $U^{(2)}$ does not vanish,
the potential $U$ becomes to have the same functional form as $U^{(1)}$
at the leading order.
%and affects the scattering of the solitons.
%In the present work, however, we do not discuss the details of
%the scattering of noncommutative solitons.
So the force between two solitons will be attractive or repulsive,
according to the sign of the coefficient.
Thus,
even if $1/V^{(2)}(\lambda)+1/V^{(2)}(0)=0$,
unless the further condition
\begin{equation}
V^{(3)}(\lambda)+V^{(3)}(0)=-{6\cdot 2!\over\lambda}V^{(2)}(0)
\label{eq:conditionv3}
\end{equation}
is satisfied in order that $U^{(2)}$ vanishes,
any static multi-soliton solution does not exist at finite $\theta$.
For instance, in a case of $\phi^3$-theory with a stable vacuum,
we have $V(\phi)=c(\phi^3-{3\over2}\lambda\phi^2)$ ($c<0$).
This satisfies $v=w$ but (\ref{eq:conditionv3}).
Then
\begin{equation}
U^{(2)}=-4\cdot {32\over 9c^2\lambda^2}{t^2\over\sinh^2 t}
\end{equation}
and it gives attractive force between two solitons.
On the other hand,
when the condition (\ref{eq:conditionv3}) is satisfied,
it may be determined by examining the $E_4$ term in (\ref{eq:expanded_e})
whether there is a force between solitons.

In this paper, we have studied
the noncommutative scalar soliton at finite $\theta$,
starting from the infinite $\theta$ limit and taking into account
the effect of finite $\theta$ 
in power series of $1/\theta$.
First we have explored the consistent solution
of the equation of motion up to the second order in $1/\theta$.
The allowed configuration depends on
the potential term $V(\phi)$ of the theory.
Then we have discussed the correction to the
moduli space of multi solitons up to that order.
The effect destroys the K\"ahler structure of the metric, which comes from
the Grassmannian nature of the moduli space.
It would be interesting to study the geometrical meaning  of
this finite $\theta$ correction.
Finally we have calculated the potential $U$
in the effective dynamics of two solitons
and discussed the force between solitons
at finite $\theta$.
In most cases, any static multi-soliton solution does not exist,
but there seems to be a possibility that static one is allowed
when we choose a particular potential $V(\phi)$.
%Furthermore application to the (unstable) D-brane system would be
%important, which is left for future study.
%{\bf Acknowledgement}:
%This work is supported in part by Grant-in-Aid from the
%Ministry of Education, Science, Sports and Culture of Japan, Priority
%Area 707 ``Supersymmetry and Unified Theory of Elementary Particles''.

\end{document}